\documentclass{JINST}
\usepackage{amsmath}   
\usepackage{amssymb}
\usepackage{array}

\title{Characterization of a Single Crystal Diamond Pixel Detector in a High Energy Particle Beam}

\author{M.~Mathes$^a$, M.~Cristinziani$^a$, H.~Kagan$^b$, S.~Smith$^b$, W.~Trischuk$^c$, J.~Velthuis$^{a,d}$, N.~Wermes$^a$\thanks{Corresponding
author}\\
\llap{$^a$}Bonn University, \\
Physikalisches Institut, Nussallee 12, D-53115 Bonn, Germany\\
\llap{$^b$}The Ohio State University, \\
Department of Physics, 191 W. Woodruff Ave., Columbus, Ohio 43210, U.S.A\\
\llap{$^c$}University of Toronto, \\
Department of Physics, 60 St. George Street, Toronto, ON, Canada \\
\llap{$^d$}now at University of Bristol,\\
Tyndall Avenue, Bristol BS8 1TL, UK\\
E-mail: \email{wermes@uni-bonn.de}
\thanks{Work supported by the German Ministerium f{\"u}r Bildung, Wissenschaft, Forschung und Technologie (BMBF) under contract no.~06 HA6PD1 and by US Department of Energy grant DE-FG02-91ER40690.
}
}

\abstract{
Diamond has been developed as a material for the detection of charged particles by ionization. Its radiation hardness makes it an attractive material for detectors operated in a harsh radiation environment e.g. close to a particle beam as is the case for beam monitoring and for pixel vertex detectors. Poly-crystalline chemical vapor deposition (CVD) diamond has been studied as strip and pixel detectors so far. We report on a first-time characterization of a single-crystal diamond pixel detector in a 100 GeV particle beam at CERN. The detectors are made from irregularly shaped single crystal sensors, 395$\mu$m thick, mated by bump bonding to a front-end readout IC as used in the ATLAS pixel detector with pixel sizes of 50 $\times$ 400 $\mu$m$^2$.
The diamond sensors show excellent charge collection properties: full collection over the entire detector volume, clean and narrow signal charge distributions with a S/N value of $\sim$100 and a hit detection efficiency of {\bf ($99.9 \pm 0.1) \%$}.
The measured spatial resolution for particles under normal incidence in the shorter pixel direction is (8.9 $\pm$ 0.1) $\mu$m.
}


\begin{document}

\section{Introduction}
The innermost tracking devices of the LHC detectors ATLAS~\cite{ATLAS-paper_2008} and CMS~\cite{CMS-paper_2008} have to cope with very intense particle radiation. Fluence levels of up to 10$^{15}$ particles per cm$^{2}$ are projected for the innermost pixel layers during the LHC lifetime~\cite{ATLAS-paper_2008}. Such large particle fluences cause damage to standard silicon pixel sensors, most dominantly by: (a) the trapping of moving charges which reduces the signal, (b) an increase in the effective doping concentration N$_{\rm eff}$ which changes the space charge in the detector substrate material, and (c) an increase of detector leakage currents. The LHC semiconductor tracking detectors have been designed and built to stand these challenges for the LHC life time~\cite{ATLAS-paper_2008,CMS-paper_2008,ATLAS-pixel-paper_2008,CMS-pixel-paper_2008}.
Plans for an upgrade of the LHC luminosity by an order of magnitude to the Super-LHC (sLHC) are already underway and with it the projected
radiation levels will rise considerably. As a consequence new types of sensors for pixel detectors need to be developed which are capable of
coping with the expected radiation fluence to 10$^{16}$ particles per cm$^{2}$. For such radiation levels standard planar silicon pixel detectors may not be suited. Poly-crystalline diamond sensors, produced by chemical vapor deposition (CVD), have been exposed to radiation levels of 1.8$\times$ 10$^{16}$ protons per cm$^2$~\cite{RD42-LHCC-2008} with the conclusion that, with the measured S/N figures and the ability to run at comparatively (to Si) low threshold settings, diamond is an attractive sensor material for large radiation fluences. Single-crystal samples have recently been shown to follow the same damage curve as poly-material~\cite{RD42-LHCC-2008}. Hence diamond, in either mono- or poly-crystalline form, is an interesting material for tracking detectors operating close to the interaction point or in radiation harsh environments close to a beam line. Other benefits of diamond as sensor material are (a) its small dielectric constant ($\epsilon_C$ = 5.7, $\epsilon_{Si}$ = 11.9) which results in smaller input capacitances per area than for silicon, and (b) its complete absence of leakage currents, which is due to its large band gap (5.5 eV). Both of these facts lead to lower noise figures than for comparable silicon detectors. Electron and hole mobilities in CVD-diamond are large (1800 cm$^2$/Vs and 1600 cm$^2$/Vs, respectively) leading to fast collection times. In addition, diamond is an excellent thermal conductor, a property which can possibly be exploited for intelligent integrated cooling concepts of pixel vertex detectors.

While poly-crystalline CVD (pCVD) diamond can be grown in wafer scale sizes~\cite{kagan_pixel2005,RD42-LHCC-2008} and large pixel detector modules have been built~\cite{kagan_pixel2005}, the grain structure of the pCVD growth process causes effects which hamper a homogeneous and smooth charge collection in comparison to depleted silicon substrates. Trapping and recombination centers limit the carrier life time and hence the distance over which moving charges can induce a signal on the pixel electrodes. In addition,  horizontal polarization fields, superimposed on the external drift field, can be generated by charges trapped at the grain boundaries. Nevertheless, charge collection distances above 300$\mu$m have been measured in pCVD diamond sensors~\cite{RD42-LHCC-2007}.

Recent advances in the production of diamonds by chemical vapor deposition has lead to single-crystal CVD (scCVD) diamond sensors with sizes larger than 1~cm$^2$~\cite{RD42-LHCC-2007} and thicknesses of up to $\sim$700~$\mu$m~\cite{RD42-LHCC-2007}. The production process is still in its development and the sensors often come in irregular shapes. scCVD detector material is believed to be superior in terms of charge collection and homogeneity than poly-crystalline diamond sensors.

Poly-crystalline diamond pixel detectors have been constructed by means of using the bump bonding and flip chip techniques as single chip devices and also as multi -- chip module detectors\cite{RD42-firstdiamond,kagan_pixel2005}; they have been tested in particle beams\cite{kagan_pixel2005,RD42-LHCC-2007}.
In this paper we report on a first-time characterization of scCVD pixel sensors using a high energy particle beam.

\section{Single crystal diamond pixel detector}\label{sCVD_device}
Single crystal diamond is fabricated using a CVD process on a single crystal substrate~\cite{CVD-ref}. Currently larger scCVD structures often come in irregular shapes of a few cm$^2$. Their charge collection properties have been measured in~\cite{RD42-LHCC-2008} showing full charge collection
for thicknesses as large as 800~$\mu$m with biasing fields below 0.1 V/$\mu$m~\cite{RD42-LHCC-2007}.
For unirradiated devices, no effect on the charge collection by pumping has been observed~~\cite{RD42-LHCC-2007} indicating that trapping of charges plays no or only a minor role in scCVD detector devices, in contrast to pCVD devices.


Our single-crystal diamond pixel detector (sensor plus electronics chip) is composed
of an irregularly shaped scCVD diamond sensor with a homogeneous thickness
of 395~$\mu$m. Both sides are metallized (Ti/W). One side is segmented into
a 50$\times$400$\mu$m$^2$ pixel pattern to match the pattern of the ATLAS pixel frontend chip (FE-I3~\cite{FEI3}).
Solder (PbSn) bumping and
flip-chipping technology at IZM (Berlin)~\cite{IZM1,IZM2} has been
employed to mate both parts. Figure~\ref{sCVD_device}(a)
shows the assembly after bump bonding. Due to the irregular shape of
the sensor only about 2200 of the 2880 possible pixels of the FE-I3 chip
are covered by a corresponding sensor cell.
\begin{figure}[thb]
\begin{center}
\includegraphics[width=0.9\textwidth]{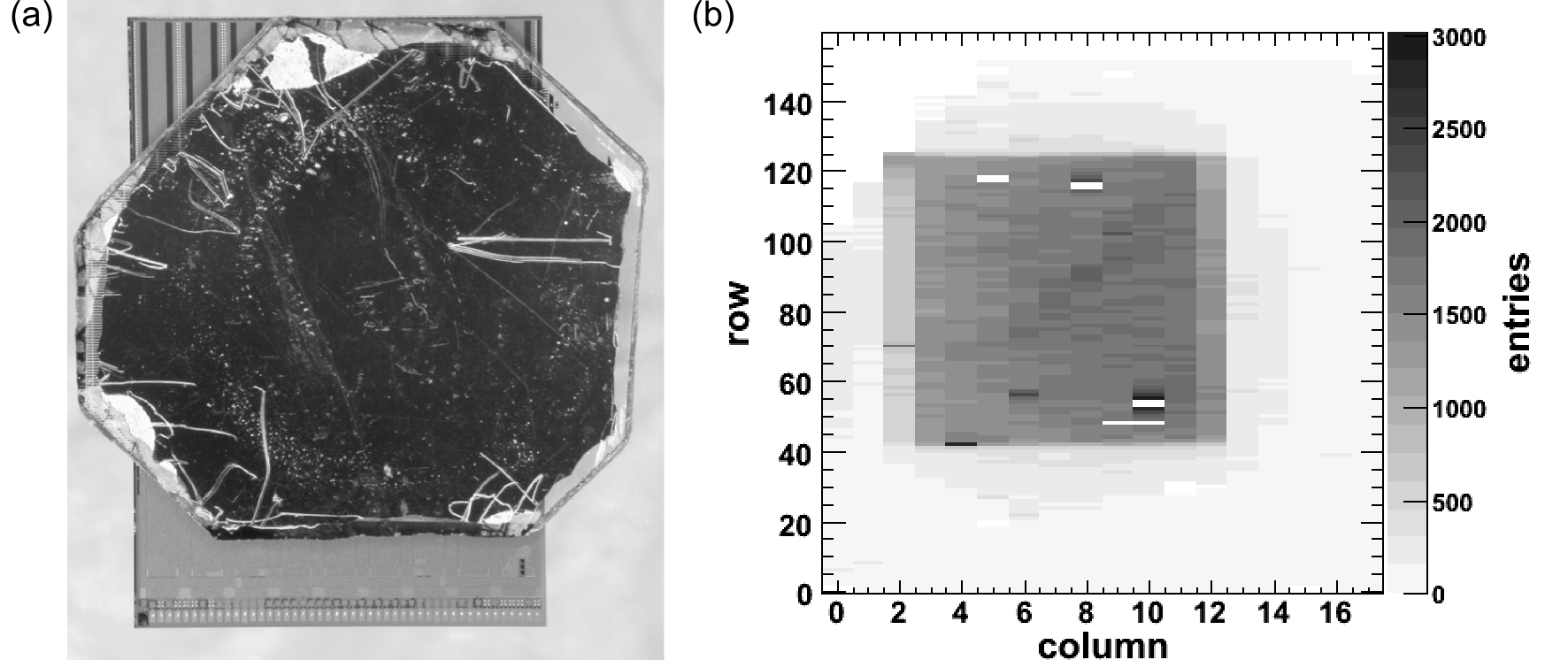}
\end{center}
\caption[]{\label{sCVD_device} (a) Photograph of the tested single crystal pixel detector. The sensor has an irregular shape. It is mated to the ATLAS FE-I3 chip. (b) Hitmap of the tested scCVD device in the testbeam. The rectangular area in the center represents the beam shadow of the scintillation trigger counters. The upper left corner of the chip is not covered by the scCVD sensor as shown in (a).}
\end{figure}

\begin{figure}[thb]
\begin{center}
\includegraphics[width=0.9\textwidth]{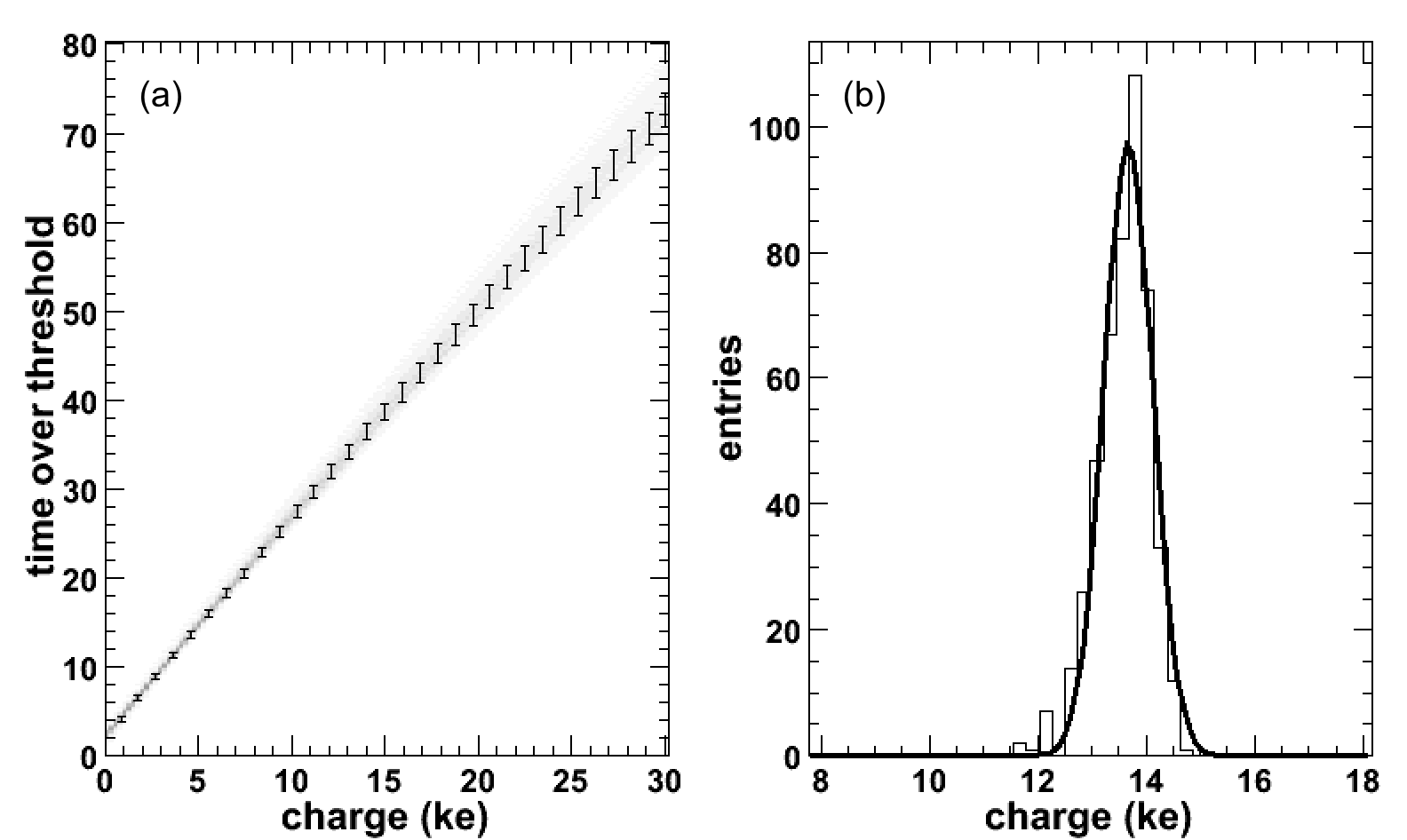}
\caption[]{(a) ToT calibration for the tested device. Shown is the ToT-output (1 ToT unit = 25 ns) as a function of the pulsed input charge. The fitted calibration curves of all 2200 pixels are superimposed. The shaded band displays the area of all calibration curves superimposed, while the error bars are the rms values of the curves at the calibration points. (b) Distribution of the charge corresponding to a fixed ToT value of 34 for one individual pixel.}
\label{ToT-calibration}
\end{center}
\end{figure}

In this paper we particularly study the response of the scCVD diamond pixel
device to high energy particles in terms of charge collection, spatial resolution and
response homogeneity. The ATLAS pixel electronics provides zero
suppression in the readout with a typical threshold level of 3000 - 4000~e$^-$ and a noise of $\sim$160~e$^-$~\cite{ATLAS-pixel-paper_2008} when used with silicon detectors.
As diamond sensors are basically free of leakage currents and have lower input capacitance, the sensor was operated at thresholds of
about 1700e$^-$ with a noise level of about 130 e$^-$, as determined from fitting the threshold curves of the pixels (s-curves). Analog information is obtained via the in-pixel measurement of
time-over-threshold (ToT) with an approximate resolution of 8~bit~\cite{FEI3}.
The ToT-calibration is obtained for every pixel individually, by fitting the ToT response to input pulses injected over a capacitance of 41.4~fF.
Figure~\ref{ToT-calibration}(a) displays the calibration curves superimposed for all pixels. The band is the area of all calibration curves superimposed, the error bars are the rms spread of the calibration at a given charge.
In fig.~\ref{ToT-calibration}(b) the distribution of the input charge for a fixed ToT value of one pixel is shown. The mean value is 13700, the fitted $\sigma$ is 500 e$^-$. Note that it is larger than the noise value of 136 e$^-$, because the ToT measurement receives contributions from leading and trailing edge uncertainties.
The calibration uncertainty is estimated to be in the range of 5 -- 10~$\%$.
This calibration is used for the determination of the charge measurement in the tested devices.

\section{Test beam setup at the CERN SPS}\label{testbeam}
\begin{figure}[thb]
\begin{center}
\includegraphics[width=0.8\textwidth]{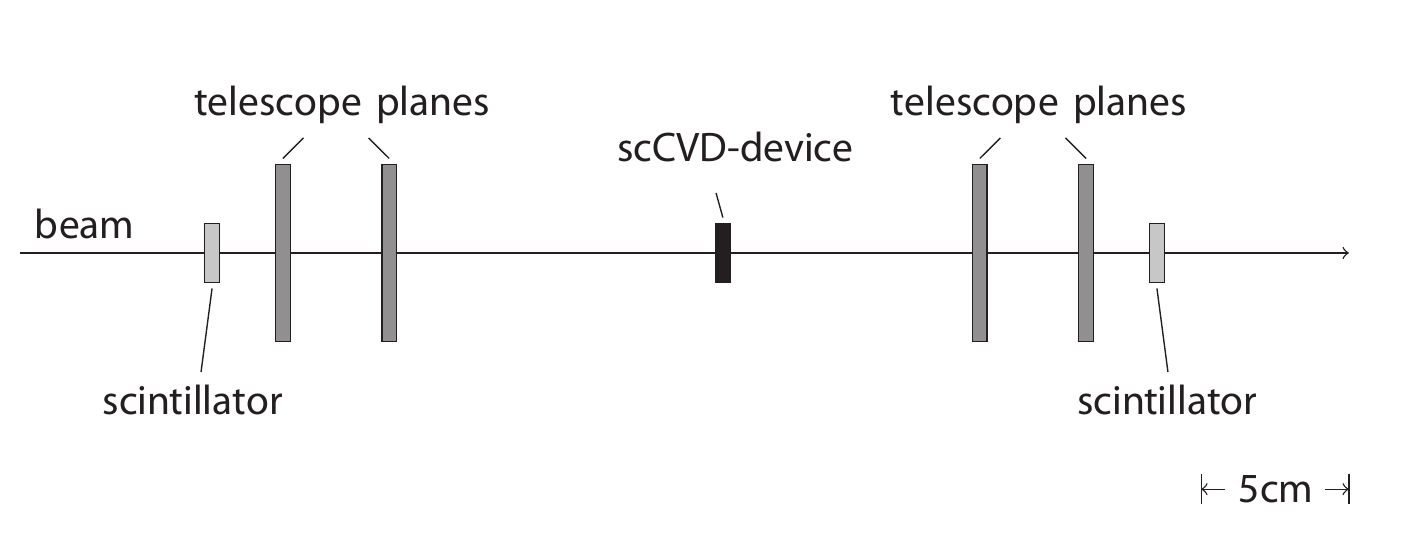}
\caption[]{Reference telescope and position of the tested scCVD pixel detector in the CERN 100 GeV pion test beam.}
\label{testbeam-setup}
\end{center}
\end{figure}

The scCVD diamond device has been tested in a 100 GeV pion beam at the CERN SPS. The setup is shown schematically in fig.~\ref{testbeam-setup}. The device under test (DUT, the scCVD detector) was placed in between two pairs of silicon microstrip
detectors~\cite{beam_telescope}. The microstrip telescope was developed for
ATLAS and consists of double sided silicon microstrip detectors with
50$\mu$m pitch strips on both sides rotated by 90$^\circ$ with respect to each other.
The S/N for the p-side, measuring the DUT x-direction, was 37 and the S/N for the n-side, measuring the DUT y-direction, was 22. Hits are read out zero
suppressed and events waiting for read out can be buffered for some
time~\cite{beam_telescope}. The latter option was, however, not
used. The telescope- and DUT-planes are triggered by the coincidence signal of two
scintillators, placed in front and behind the setup. The precision obtained in
the plane of the DUT is about 5~$\mu$m in both spatial
directions.
The beam divergence is measured to be less than 0.2~mrad. The beam incidence angle with respect to the DUT plane
is at most 1$^\circ$.
The average data taking rate was limited by the readout system to
around 50-60 Hz.

Only tracks from events with a single hit in each of the telescope
planes were selected for the characterization measurements.
Hits in the scCVD device were accepted if the seed position
of the cluster is found in the predicted pixel or in a neighboring or next to neighboring (for the short pixel dimension) pixel.
Otherwise it is counted as a miss in the detector efficiency calculation.

Figure~\ref{sCVD_device}(b) shows the hitmap in the plane of the characterized scCVD device in the test beam. The upper left corner, in which the FE-chip is not covered by the sensor is clearly visible as well as the shadow of the scintillation trigger counters illuminating the DUT with an almost quadratic profile. Two
pixels had no bump contact and less than 20 were masked off during operation because they were noisy or impossible to tune to the low threshold setting.

%

\section{Results}\label{results}
\begin{figure}[thb]
\begin{center}
\includegraphics[width=0.8\textwidth]{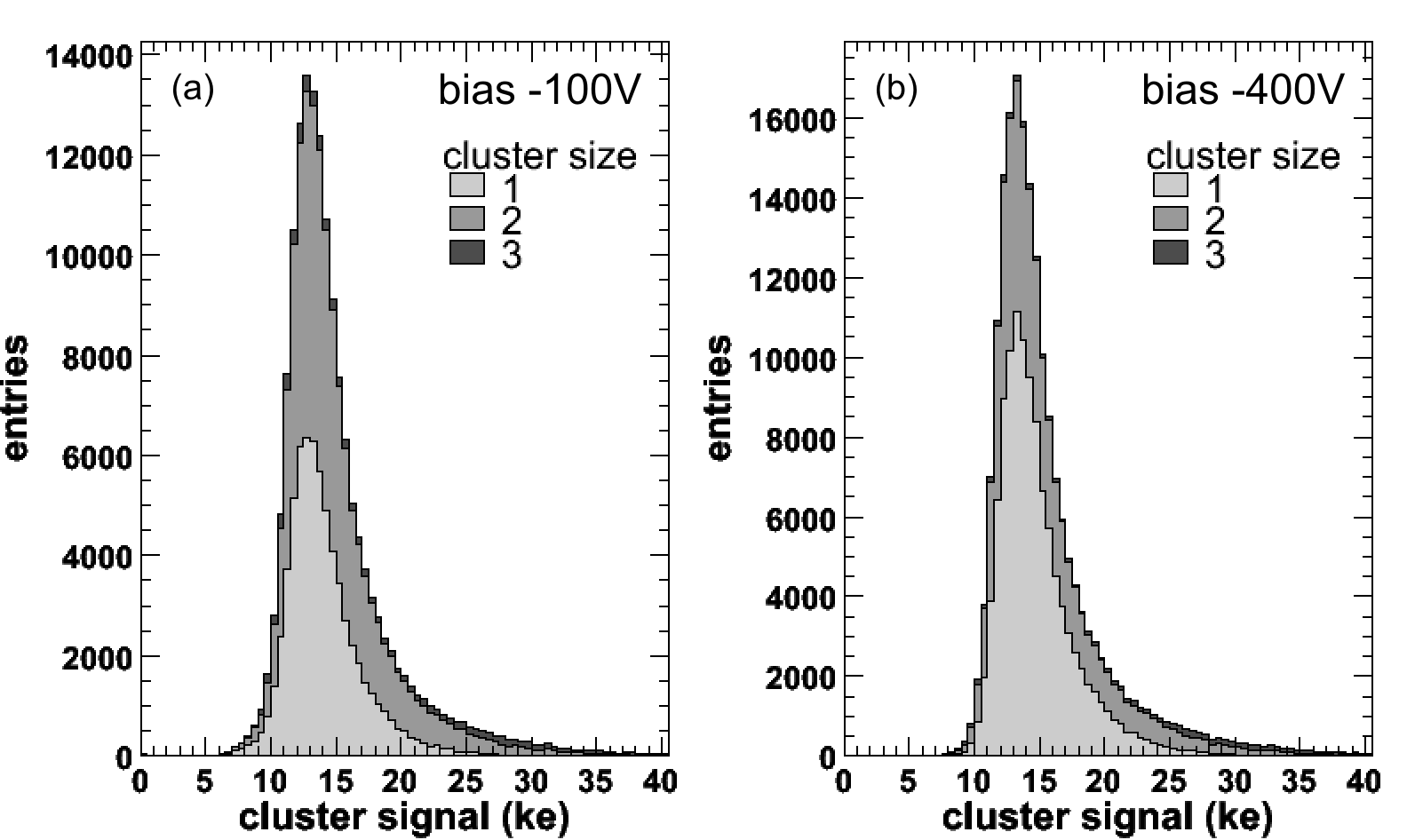}
\caption[]{Measured cluster charge distributions for bias voltages of (a) 100 V and (b) 400 V. The contributions of one, two and three pixel clusters are indicated by the shaded areas.}
\label{landau}
\end{center}
\end{figure}

\paragraph*{Charge collection and cluster size}
The diamond sensor can be operated with a fairly large range of biasing voltages. Unlike in silicon the electric field inside the sensor is fairly uniform
for good crystals~\cite{TCT-Fink}. Over the area that could be investigated (i.e. covered by beam and electronics) the response of the detector was uniform for bias voltages larger than 100 V (0.25 V/$\mu$m). As the drift velocity is directly proportional to the field, higher fields cause faster charge collection and hence a smaller spread of the charge cloud by diffusion. This is illustrated in fig.~\ref{landau} where the measured charge distributions are plotted for fields of roughly 0.25 V/$\mu$m in (a) and 1V/$\mu$m in (b), respectively.

For high energy pions of 100 GeV, the mean energy loss in diamond due to ionization is about 10$\%$ larger than for particles in the minimum of the Bethe-Bloch curve~\cite{PDT}. Also the ratio between mean and most probable value of the charge distribution (Landau distribution) is thickness dependent and differs from that observed in silicon for the same thickness. For thicknesses between 200$\mu$m and 400$\mu$m the distribution for diamond is narrower than for silicon by 5--10$\%$. To estimate the most probable energy loss and charge deposit in our device and compare it with the calibrated charge measurement, we have followed\cite{PDT} and the detailed descriptions given in~\cite{bichsel_1988}. Attempted calibration cross referencing by using standard radioactive sources for silicon detectors (e.g. $^{241}$Am or $^{57}$Co) turned out to be very difficult with the limited ToT charge resolution due to the much larger Compton contribution to photon absorption in diamond in comparison to silicon and the appearance of several nearby emission lines in many of these sources.

\begin{figure}[thb]
\begin{center}
\includegraphics[width=0.9\textwidth]{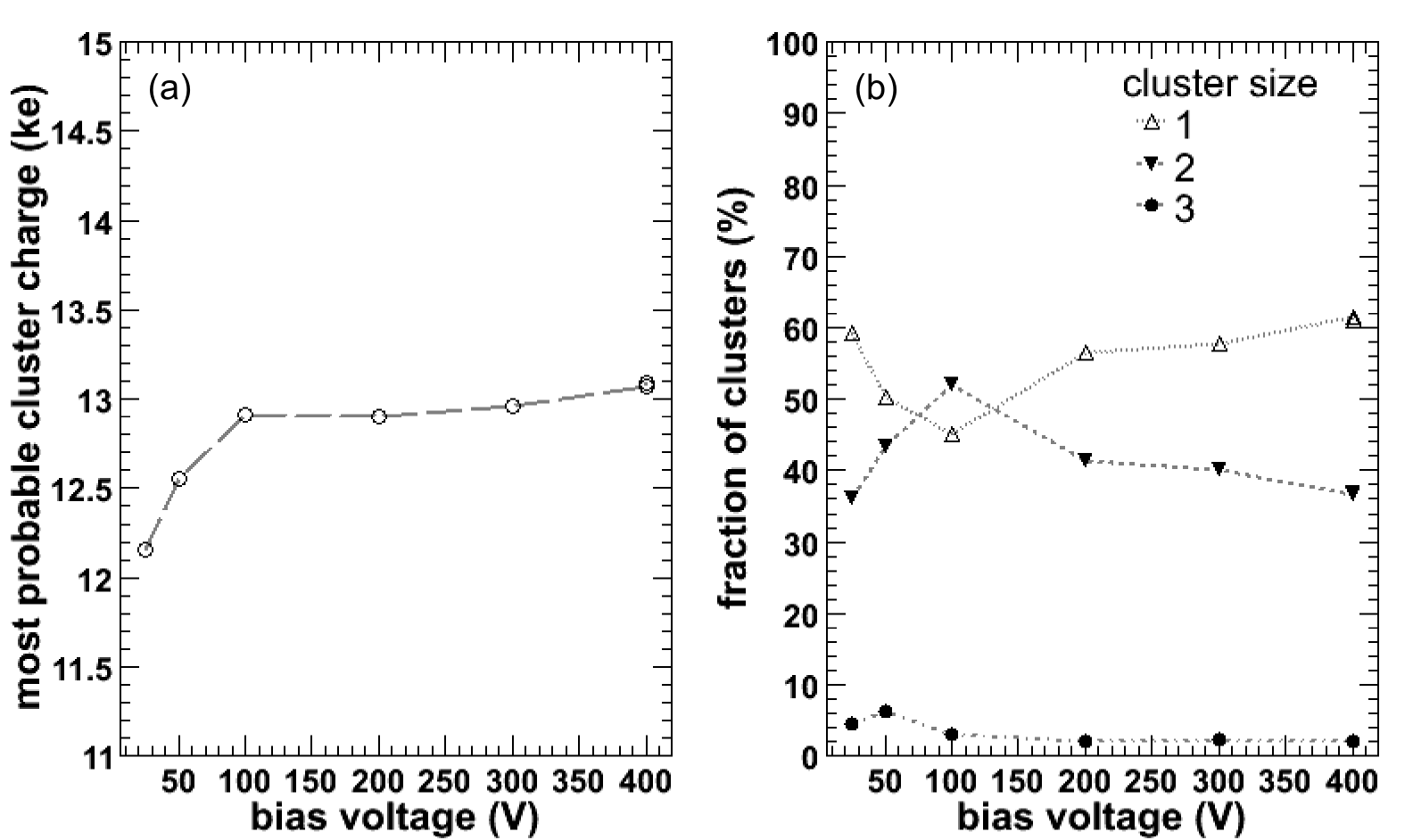}
\caption[]{Most probable values of the charge distribution (a) and fractions of charge cluster sizes (b) as a function of the bias voltage. }

\label{charge-Udep}
\end{center}
\end{figure}

We adopt the information given in~\cite{bichsel_1988} to estimate the charge deposited in our scCVD diamond sensor by 100 GeV pions.
For a thickness of 395~$\mu$m the mean energy loss in carbon for a minimum ionizing particle is 1.725 MeV cm$^2$/g. With the density for diamond of 3.52 g/cm$^3$ and a mean energy for e/h creation of 13.1 eV, we find the most probable value (MPV) of the charge distribution by scaling the dependence on the thickness as given in \cite{bichsel_1988} for silicon by the respective density ratio. Hence we expect a most probable charge value in the scCVD sensor of 13900 e$^-$. The measured value, using the ToT to charge calibration outlined above, yields 13100 e$^-$ at 400~V bias voltage (fig.~\ref{landau}(b)), in fair agreement with the expectation within the calibration uncertainty (5 -- 10$\%$). We note also that for ATLAS silicon pixel modules with the same readout chip it has been observed~\cite{stockmanns} that the absolute calibration
to the $^{241}$Am 59.6 keV line in Si came out low by 8$\%$ which indicates that the charge injection capacitance may deviate from its design value by this amount. However, other explanations are conceivable. If one corrects the measured peak value for this mismatch the agreement with the expectation is within a few percent. The full width at half maximum in fig.~\ref{landau}(b) is 4000 e$^-$, 30$\%$ of the MPV. The theoretical value following~\cite{bichsel_1988} is 4200 e$^-$.
With different shadings also the contributions from different cluster sizes to the Landau-type shape are indicated in fig.~\ref{landau}.
The spatial charge distribution (charge cloud) becomes narrower for higher voltages, resulting also in a larger fraction of single pixel clusters at higher bias voltages.
Figure~\ref{charge-Udep}(a) shows the measured MPV as a function of bias voltage. The curve saturates for bias voltages above about 100 V, indicating that for this single crystal diamond device the charge collection is complete. A small increase towards high biasing voltages is still observed and expected since for higher
fields the probability of single pixels clusters increases rendering charge losses below neighbor pixel thresholds less likely.

Figure~\ref{charge-Udep}(b) shows the fractional decomposition of clusters (in Y direction) as a function of the bias voltage, showing that for full charge collection and low thresholds single and double hit clusters have roughly equal sharing (60:40).

The overall efficiency to find a hit near a track point (i.e. within 5x3 pixels, asymmetric due to the rectangular pixel size) extrapolated on the plane of the DUT is found to be $99.9 \% \pm 0.1 \%$.

\begin{figure}[thb]
\begin{center}
\includegraphics[width=0.8\textwidth]{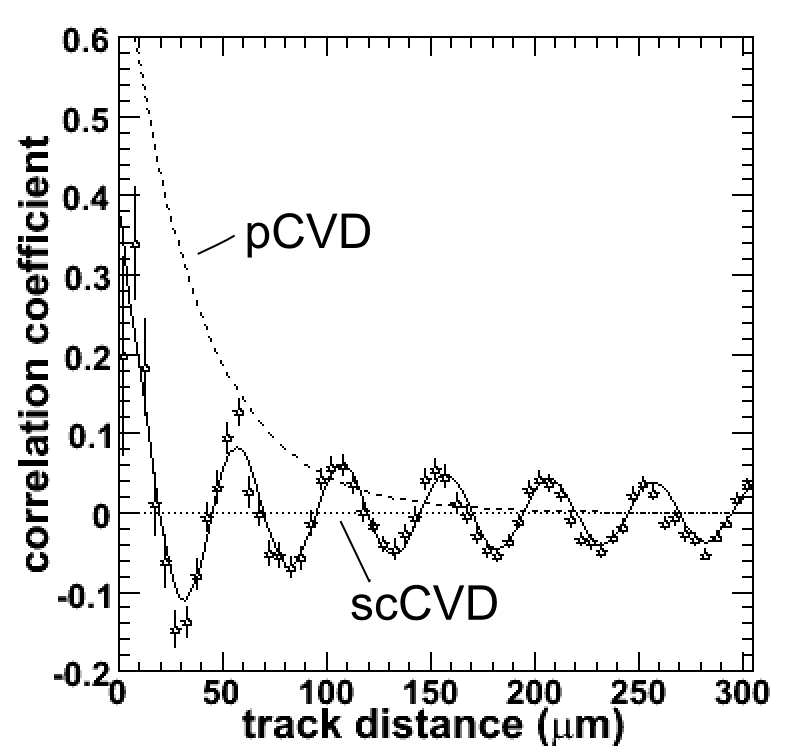}
\caption[]{Correlation coefficient between the spatial residuals of two events, as a function of the separation between the tracks of the two events. The solid line is a fit with a function containing a sin(ax)/$\sqrt{x}$ plus an exponential term. The dashed line displays the correlation found in~\cite{lari05} for a pCVD diamond device showing a pronounced exponential contribution. The dotted line is the exponential contribution for the device studied in this paper.}
\label{residual-correlations}
\end{center}
\end{figure}

\paragraph*{Space point reconstruction}
In previous studies of poly-crystalline diamond~\cite{lari05}, the
pCVD grain structure and the trapping of electric charges at their
boundaries which may cause horizontal polarization fields, were
identified as one explanation that affects the charge collection and
thus the space point reconstruction in pCVD devices. As in \cite{lari05} we plot in
fig.~\ref{residual-correlations} the correlation of the reconstructed
Y coordinates (corresponding to the short 50 $\mu$m pixel dimension)
of the measured hits of two nearby tracks taken in different events,
as a function of the distance of the tracks' impact points (as
determined from the telescope extrapolation). The sinusoidal shape of
this distribution reflects the pixel pitch of 50 $\mu$m, damped by a
radial distance effect. The data are well fitted by a functional form
$a \, $sin$(bx)/\sqrt{x} + c\,$ exp$(-x/r_c)$, with $c$ being
consistent with zero. In~\cite{lari05}, albeit with a rather poor
quality pCVD sensor compared to present standards, it was shown that
in comparison with silicon, pCVD diamond -- as a consequence of the
pCVD grain structure -- showed a net correlation developing for small
two track distances and a correlation length $r_c = $ 36 $\mu$m for
the exponential term was measured. This effect, observed in
\cite{lari05}, is indicated by the dashed line in
fig.~\ref{residual-correlations}.
The measurements presented here for 400 V bias voltage now demonstrate
that for single crystal diamond such a net correlation no longer
exists, i.e. no evidence for horizontal polarization fields are
observed.

\begin{figure}[thb]
\begin{center}
\includegraphics[width=0.75\textwidth]{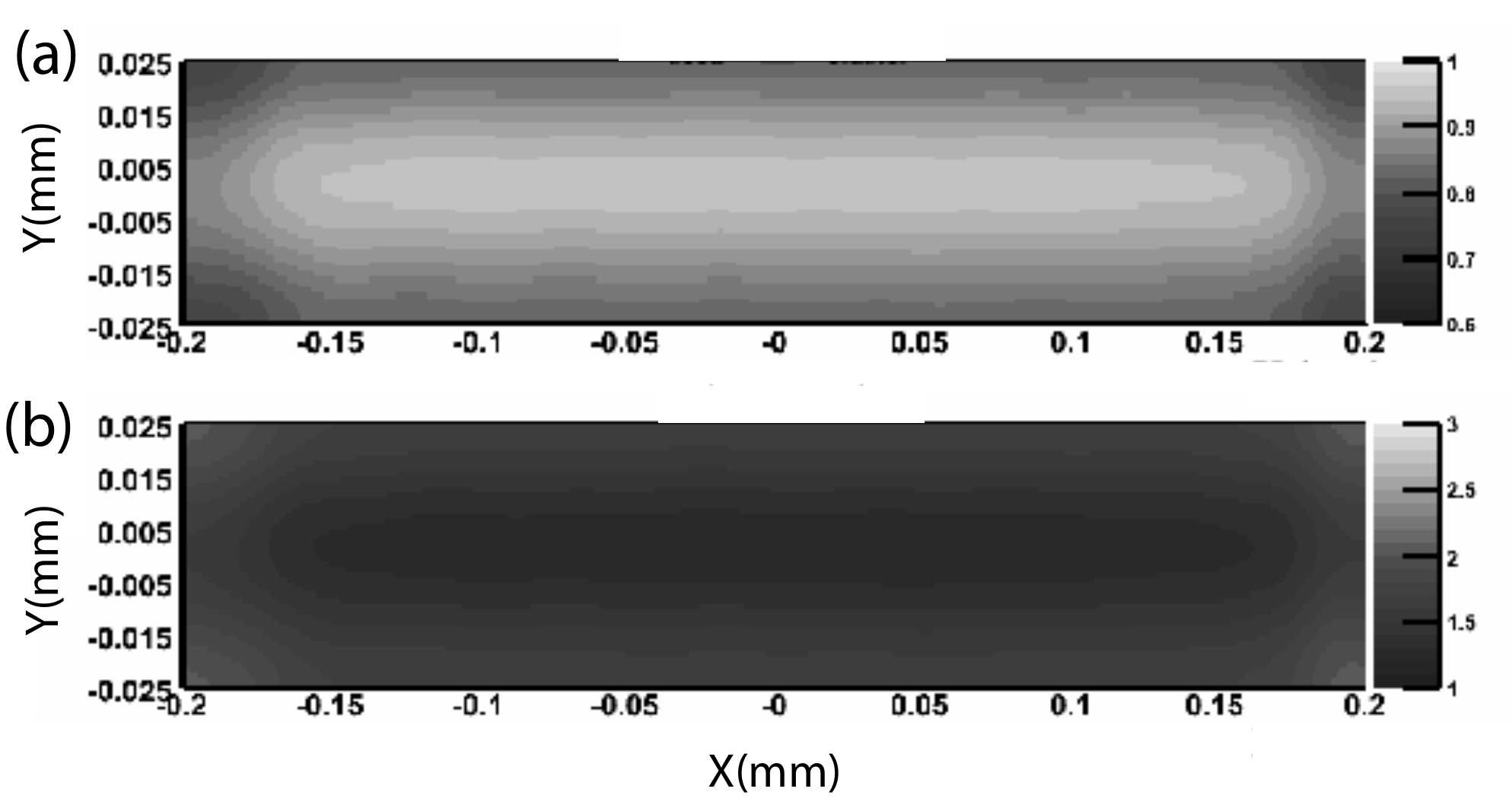}
\caption[]{Maps over the area of one pixel of (a) the fractional charge seen in the pixel with the largest signal (seed
pixel), and (b) the cluster size (number of pixels above threshold) for normal incident particles.}
\label{charge-sharing}
\end{center}
\end{figure}
We show in fig.~\ref{charge-sharing} that proper sharing of signal charge is observed in scCVD diamond pixel sensors as it is in silicon sensors.
In fig.~\ref{charge-sharing}(a) the fractional signal charge in the pixel with the largest charge of a hit-cluster (seed pixel) is shown as a map over the pixel area. Fig.~\ref{charge-sharing}(b) is the corresponding map of the cluster size (number of pixels above threshold).
\begin{figure}[thb]
\begin{center}
\includegraphics[width=0.9\textwidth]{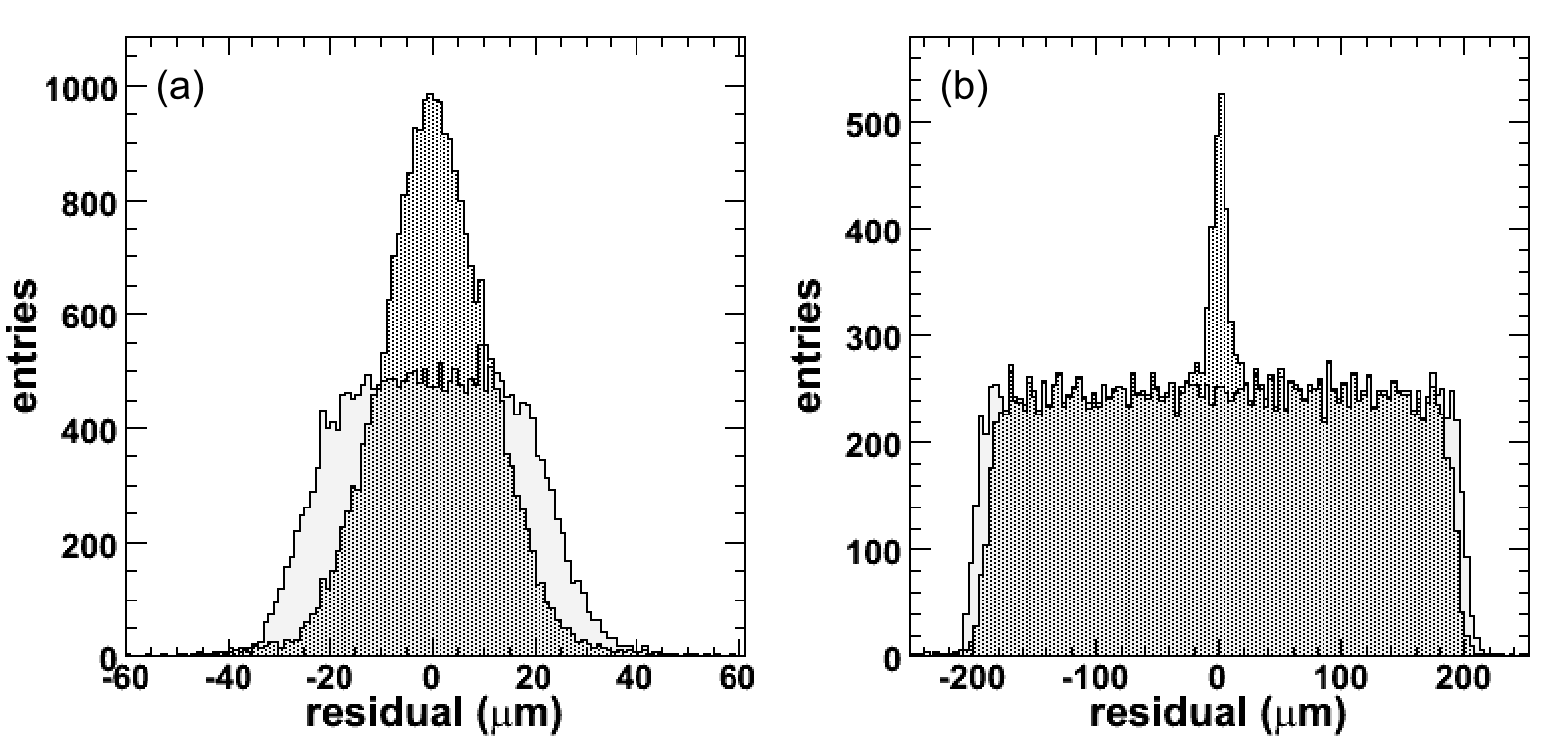}
\caption[]{Spatial resolution of the scCVD device measured with
respect to the reference telescope, using digital information only (light shaded histograms)
and analog information via the $\eta$-algorithm~\cite{eta-algorithm} (dark histograms).
Bias voltage was 200 V. Plotted is the difference between the telescope space point and the measured space
point. (a) resolution in the 50$\mu$m direction, (b) resolution in
the 400$\mu$m direction.}
\label{resolution}
\end{center}
\end{figure}

Finally, the spatial resolution is measured by plotting in
Fig.~\ref{resolution} the difference between the track position
predicted by the telescope on the plane of the DUT (the scCVD sensor)
and the reconstructed hit of the DUT device. Figure~\ref{resolution}
shows this distribution for both directions of the pixel. For the hit
reconstruction first only the digital information is used (light gray
distribution), i.e. the pixel with the largest charge signal above threshold
collected in a cone around the extrapolated track position is taken as
the hit pixel and its center is assumed to be the reconstructed
position. The expected digital resolutions of pixel pitch divided by
$\sqrt{12}$ in both directions, i.e. 14.4~$\mu$m and 115.5 $\mu$m,
respectively, folded by the resolution of the track extrapolation of
about 5$\mu$m, is observed. Position reconstruction which exploits the
analog information available via the ToT readout is done by using the
$\eta$-algorithm~\cite{eta-algorithm}. This algorithm provides an improved resolution for clusters with more than one hit, especially double-hit clusters.
Single-hit clusters effectively give binary response using $\eta$.
The corresponding residual distribution is shown by the
dark histograms in fig.~\ref{resolution}. It shows a pronounced peak with good resolution for clusters with two pixel hits, for which the $\eta$-algorithm improves the resolution substantially. In the short pixel direction (Fig.\ref{resolution}(a)), where the fraction of two hit clusters is
about 40$\%$ the intrinsic resolution of the device, after quadratically subtracting the telescope
extrapolation error of 5\,$\mu$m, becomes (8.9 $\pm$ 0.1)\,$\mu$m for a bias
voltage of 200 V. The error quoted is obtained from the fit to the residual distribution. Systematic errors are not included as their exact determination turned out to be difficult, involving spatial S/N variations, alignment precision and its variation, as well as data sample (i.e. time/run) dependencies. From repeated analysis
of several data samples, we conclude that an additional uncertainty in the order of 0.5\,$\mu$m has to be assumed.

Selecting events with two hit clusters only improves the resolution to $\sim$5$\mu$m, telescope resolution quadratically subtracted, with the same uncertainty as quoted above. In the direction of the long pixel dimension (Fig.\ref{resolution}(b)), the resolution for events with two-hit clusters ($\sim$10$\%$) is an order of magnitude better than for events with single hit clusters.
\begin{figure}[h!]
\begin{center}
\includegraphics[width=0.65\textwidth]{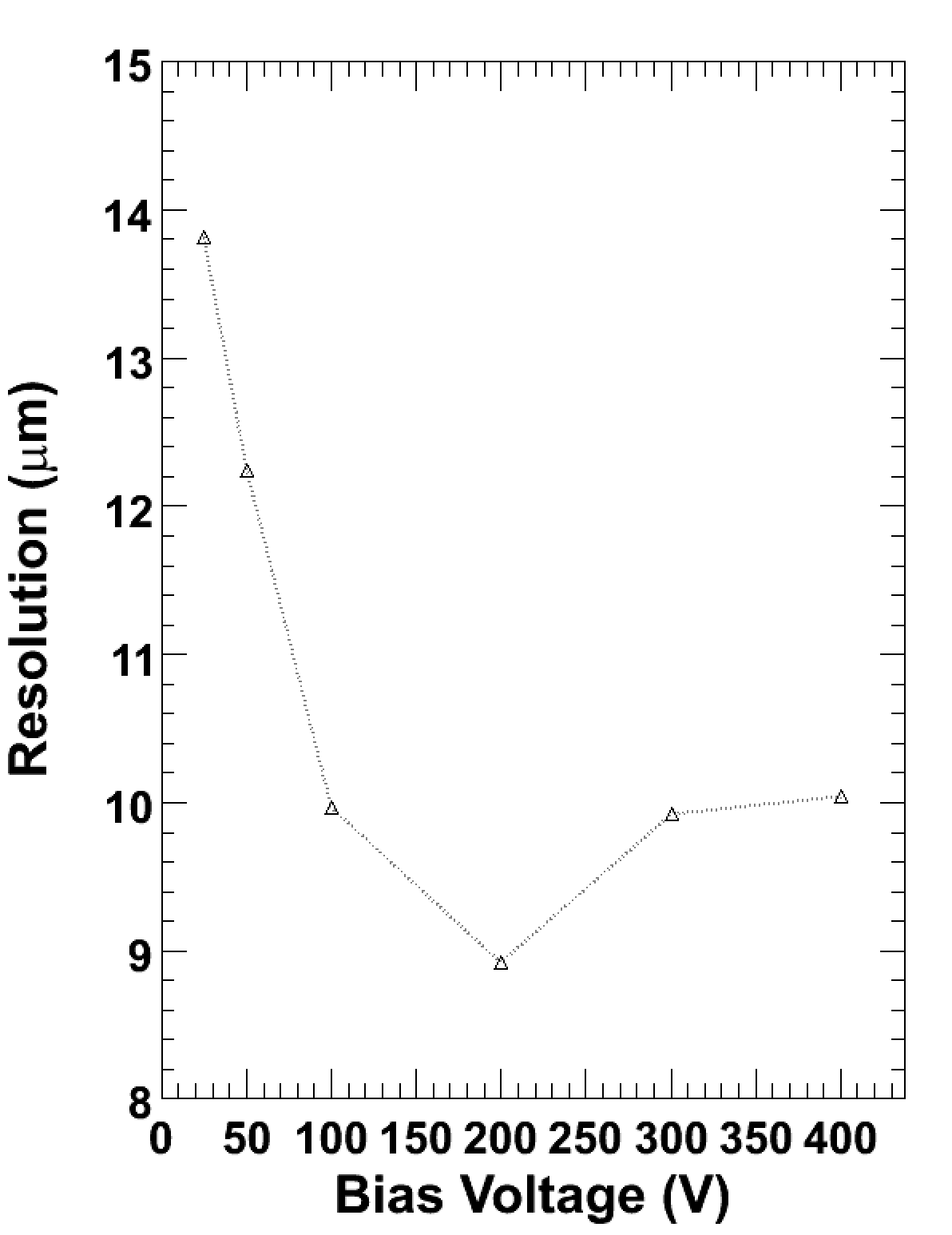}
\caption[]{Spatial resolution as a function if the applied bias voltage.
The telescope uncertainty is subtracted. The errors (not shown) are in the order of 0.1$\mu$m.}
\label{resVsV}
\end{center}
\end{figure}

Figure~\ref{resVsV} shows that the space resolution depends on the bias voltage. The charge collection efficiency and hence the total collected charge are field dependent until full charge collection is obtained. Also the sharing of charge between pixels due to diffusion depends on the field inside the sensor. The measured  resolution shows an optimum between both effects at bias voltages of 200~V (fig.~~\ref{resVsV}). The contribution of the telescope resolution is again subtracted in this figure. The uncertainty on the resolution measurements is in the order of 0.1$\mu$m.

\section{Conclusions}
A single crystal chemical vapor deposition (scCVD) diamond sensor has been assembled as a hybrid pixel detector by using bumping and flip-chip technology with the ATLAS pixel readout chip FE-I3. As a pixel detector, scCVD diamond has been characterized for high energy particle detection for the first time. A high energy (100 GeV) pion test beam with a 4-plane microstrip beam telescope as a reference system has been used. The detector properties of scCVD diamond are free of effects observed previously with poly-crystalline (pCVD) devices: full charge collection over the sensor thickness ($\sim$400$\mu$m) for bias
voltages above 100V (E=0.25V/$\mu$m) is observed, no evidence for charge trapping and polarization fields is seen. The observed signal charge is $\sim$13000 e$^-$. The operation of the device is characterized by low noise values (130~e$^-$) and low threshold (1700 e$^-$) settings. The measured charge distribution is a narrow Landau distribution due to the high density of the material.
Hit detection efficiencies larger than 99.9$\%$ have been measured. The spatial resolution, obtained by using the digital hit information is consistent with that expected for a binary resolution. Exploiting charge sharing in the short pixel direction yields a measured intrinsic resolution of (8.9 $\pm$ 0.1) $\mu$m.

\section*{Acknowledgments}
The authors would like to thank the CERN SPS staff
for their help during data taking.
%


\end{document}